\documentclass{aa}

\usepackage{natbib}
\usepackage{graphicx}

\usepackage{natbib}

\begin{document}

\title{On the probability distribution function of the mass surface density of molecular clouds I}
\titlerunning{On the PDF of the  mass surface density of molecular clouds I}

\author{J\"org Fischera}
\institute{Canadian Institute for Theoretical Astrophysics, University of Toronto, 60 St. George Street, ON M5S3H8, 
	Canada\label{inst1}
	}


\abstract{
	The probability distribution function (PDF) of the mass surface density is an essential characteristic
	of the structure of molecular clouds or the interstellar medium in general. 
	Observations of the PDF of molecular clouds indicate a composition of a broad distribution
	around the maximum and a decreasing tail at high mass surface densities. The first component is
	attributed to the random distribution of gas which is modeled using a log-normal function while
	the second component is attributed to condensed structures modeled using a simple power-law.
	The aim of this paper is to provide an analytical model of the PDF of condensed structures
	which can be used by observers to extract information about the condensations.
	The condensed structures are considered to be either spheres or cylinders with
	a truncated radial density profile at cloud radius $r_{\rm cl}$. The assumed profile is of the form
	$\rho(r)=\rho_{\rm c}/(1+(r/r_0)^2)^{n/2}$ for arbitrary power $n$
	where $\rho_{\rm c}$ and $r_0$ are the central density and the inner radius, respectively.
	An implicit function is obtained
	which either truncates (sphere) or has a pole (cylinder) at maximal mass surface density. 
	The PDF of spherical condensations and the asymptotic PDF of cylinders in the limit of infinite
	overdensity $\rho_{\rm c}/\rho(r_{\rm cl})$ flattens for steeper
	density profiles and has a power law asymptote at low and high mass surface densities
	and a well defined maximum. The power index of the asymptote $\Sigma^{-\gamma}$ of the logarithmic
	PDF ($\Sigma P(\Sigma)$) in the limit of high mass surface densities is given by $\gamma = (n+1)/(n-1)-1$ (spheres)
	or by $\gamma=n/(n-1)-1$ (cylinders in the limit of infinite overdensity).
	}


\maketitle

\section{Introduction}

{Observations of the structure of molecular clouds provide insights
about the physical processes in the cold 
dense phase of the interstellar medium and will give us a better understanding 
how they evolve and eventually form stars. They are furthermore essential to test
theoretical models of the origin of the stellar mass function
\citep{Padoan1997a,Elmegreen2001, Padoan2002, Hennebelle2008, Elmegreen2011, Hopkins2013a}
and the star formation rate \citep{Krumholz2005,Padoan2011,Hennebelle2011,Federrath2012} which are
both thought to be related to the density structure of a turbulent molecular gas.

The high resolution and sensitivity of modern telescopes allows a detailed
analysis of the 1-point statistic or probability distribution function (PDF) of the mass surface density
of molecular cloud gas. They are
obtained using either the reddening of stars \citep{Kainulainen2009,Froebrich2010,Lombardi2010,
Kainulainen2013, Alves2014}
or more recently the infrared emission of dust grains \citep{Hill2011,Hill2012,Schneider2012,Schneider2013a,Schneider2013b,Russeil2013}.

Despite the complexity of the molecular clouds the observed PDFs of
the mass surface density show very similar properties. They all are
characterized by a broad peak and a tail at high mass surface densities approximately
given by a power law.
The PDF at low mass surface densities around the broad peak is attributed to randomly
moving gas commonly referred to as 'turbulence' while the tail is attributed to 
self-gravitating cloud structures. The relative amount of the two different components 
seems to be related to the star formation activity in the cloud as discussed by \citet{Kainulainen2009}. 
While non-star forming clouds as the 'Coalsack' or the 'Lupus V' region show only a very low 
or no evidence of a tail
the PDFs of star forming clouds as 'Taurus' or 'Orion' are characterized by a strong tail with no clear 
separation between the two components. 

The observations seem to be broadly consistent with current simulations of turbulent molecular clouds. 
Turbulence would naturally create a wide range
of densities and simulations of driven isothermal turbulence have shown that the corresponding
PDF has a log-normal form \citep{Vazquez1994,Padoan1997b,Passot1998}, 
a result which has been confirmed analytically \citep{Nordlund1999}. The projection
of the density of those simulations has also been found to be closely log-normal in shape 
\citep{Ostriker2001,Vazquez2001,Federrath2010,Brunt2010c}. Deviations are expected for non isothermal
turbulence which produces higher probabilities at low or high densities \citep{Scalo1998,Passot1998,Li2003}.
More recent simulations of forced turbulence also show depending on the assumed forcing
for the PDF of the volume density a deviation from the log-normal function with enhanced probabilities at low 
densities \citep{Federrath2008b,Konstandin2012,Federrath2013b}. The functional form is as
shown by \citet{Federrath2013b} approximately described by a
statistical function proposed by \citet{Hopkins2013b}.

{Simulations of the time evolution of molecular clouds have shown that at late stage the PDF of the volume density
would develop a tail-like structure at high density values \citep{Klessen2000,Dib2005,Vazquez2008}. The
same behavior is also seen in the PDF of the mass surface density \citep{Ballesteros2011,Kritsuk2011,Federrath2013a}.



Currently, observed PDFs are analyzed using a 
log-normal function for the peak and a simple power law for the tail, respectively.
The log-normal function allows a first estimate of the density contrast of the volume density in a turbulent 
medium based on the fundamental relation of the statistical properties of the mass surface density
and the ones of the volume density as provided by \citet{Fischera2004a} and 
also by \citet{Brunt2010b,Brunt2010c}. 
The interpretation of the tail is frequently based upon a simple power law density profile $\rho(r)\propto r^{-n}$ 
of spheres where the PDF of the mass surface density is also a power law. In case of the logarithmic
 PDF ($\Sigma P(\Sigma)=P(\ln \Sigma)$)\footnote{The PDF of the logarithmic values of the mass surface density
is referred to as logarithmic PDF while the PDF of the absolute values of the mass surface density
as linear PDF.}  the corresponding power law would be
$\Sigma^{-\gamma}$ with $\gamma=2/(n-1)$ \citep{Kritsuk2011,Federrath2013a}. Applying this relation
the slope of the tail of the PDF of a number of star forming molecular clouds indicates 
a radial density profile with a power index $n\sim 2$ \citep{Schneider2013b} as expected for collapsing clouds.
A different approach has been chosen by \citet{Kainulainen2013} who also applied a log-normal function 
to the tail. 


However, the analytical functions show partly strong deviations to the observed curves.
Most of the PDFs published by \citet{Kainulainen2009} reveal a tail at low
mass surface densities below the peak which cannot be explained in terms
of the simple analytical function. The tail at high mass surface densities 
has several features which are not expected using
simple power law profiles of the radial density. 
Foremost, the tail is restricted to a certain range of mass surface densities.
For a number of PDFs published by \citet{Kainulainen2009} the tail indicates a strong cutoff 
or a strong change of the slope around $A_V=6-10~{\rm mag}$. 
The interpretation of the tail is further complicated by the observational facts that 
condensed clouds are generally located on a certain background level and that they 
are restricted to small regions within the cloud complex as e.g.
in case of the Rosette molecular cloud \citep{Schneider2012}. The tail is therefore not 
necessarily a simple power law nor directly related to the radial density profile.
Furthermore, the analytical functions do not provide a physical explanation for the
peak position of the PDF which occurs in case of a number of molecular clouds around 
$A_V\sim 1~{\rm mag}$. }

In this and the following papers an analytical {physical} model of the {global} 
PDF of molecular clouds is {developed} 
which resembles the {main} observed features {and is meant to derive basic physical parameters of star
forming molecular clouds as the pressure and the density contrast in the turbulent gas.}
This paper focuses on the 1-point statistical properties
of individual condensed structures, assumed to be spheres and cylinders.

In Sect.~2 an analytical solution of the mass surface density and the corresponding PDFs for the 
considered shapes is presented which is based on a truncated analytical density profile 
widely used in astrophysical problems. In Sect.~3 the properties of the PDFs {are} discussed and
asymptotes for low and high mass surface densities are provided. Also studied is
the location of the maximum position of the logarithmic and linear PDF.
A summary is given in Sect.~4. The technical details can be found in the appendices.
}

\section{Model of the PDF of the mass surface density of condensed structures}

\subsection{\label{sect_densityprofile}Radial density profiles}

Let us assume for the condensed structures a simple analytical density profile given by
\begin{equation}
	\label{eq_densityprofile}
	\rho(r) = \frac{\rho_{\rm c}}{(1+(r/r_{0})^2)^{n/2}},
\end{equation}
where $\rho_{\rm c}$ is the density in the cloud center and $r_{\rm 0}$ the inner radius.
The density profile has a flat part in the inner region $(r\ll r_0)$ which approaches asymptotically a power
law $\rho\propto r^{-n}$ at large radii $(r\gg r_0)$. In studies of stellar clusters this inner radius is 
frequently referred to as `core radius' \citep{King1962,King1966a,King1966b}. 
The analytical profile is used in astrophysics for its convenience and as generalization of physical density profiles
to model the stellar surface brightness of Globular clusters (e.~g. \citealt{Elson1987,Elson1992}) and more recently
the dust emission of dense filaments (e.~g. \citealt{Arzoumanian2011,Malinen2012,Juvela2012}).

We make another reasonable assumption 
that the profile is truncated at radius $r_{\rm cl}$ as expected if the clouds are 
cold structures embedded in warmer gas. In case of pressure equilibrium the gas pressure at the outer boundary
of the condensed structure would be equal to the pressure $p_{\rm ext}$ in the surrounding medium. We refer
to the inverse of the density ratio $\rho(r_{\rm cl})/\rho_{\rm c}=q$ 
of the density at cloud radius and cloud center as `overdensity'. In case of isothermal clouds this ratio
is identical with the term {'overpressure'} used to characterize the gravitational state of self-gravitating structures 
in previous studies of pressurized clouds \citep{Fischera2008,Fischera2011,Fischera2012a,Fischera2012b}.

Specific profiles with certain values of the power 
$n$ are known solutions for the physical problem of self-gravitating gaseous clouds.
The profile for $n=5$ is valid for a gaseous sphere where the temperature of the gas is regulated
by the adiabatic law with a ratio 1.2 of the two specific heats
\citep{Schuster1883, Jeans1916}. This density profile has been applied to describe the
surface brightness profiles of globular clusters and is known as 
Plummer-model \citep{Plummer1911,Plummer1915}. 
The density profiles which correspond to the power indices $n=2$, $3$, and $4$
are related to profiles of isothermal self-gravitating clouds. 

%

The profile with $n=4$ is the exact solution for a self-gravitating isothermal cylinder 
\citep{Stodolkiewicz1963,Ostriker1964}. 
The radial density profile of pressurized isothermal self-gravitating spheres, referred to
as Bonnor-Ebert sphere \citep{Ebert1955,Bonnor1956}, cannot be expressed through a 
simple analytical formula. However, 
the profile for spherical clouds up to an overdensity $\sim 100$ is
in excellent agreement with the analytical profile with $n=3$. 
The profile with $n=3$ is also identical with the well known King model without truncation
\citep{King1962,King1966a}. The radial density profile of spheres
with higher overdensity might be approximated with a profile where $n=2$. 
The statistical properties of isothermal clouds are analyzed in more detail in a forthcoming paper \citep{Fischera2014b}. 

\subsection{\label{sect_msurfprofile}Mass surface density profiles}

The   mass surface density profile of a truncated analytical density profile (Eq.~\ref{eq_densityprofile}) {of 
a sphere or a cylinder seen at inclination angle $i$ where $i=0^\circ$ refers to a cylinder seen edge-on} is given by
\begin{equation}
	\Sigma_n (r) = \frac{2}{\cos^\beta i}\int\limits_0^{r_{\rm cl}\sqrt{1-(r/r_{\rm cl})^2}} {\rm d}l
	\frac{\rho_{\rm c}}{(1+(r^2+l^2)/r_{\rm 0}^2)^{n/2}}.
\end{equation}
where $\beta=0$ for spheres and $\beta=1$ for cylinders.
In the following it is convenient to define a parameter 
\begin{equation}
	y_n = (1-q^{2/n})(1-x^2),
\end{equation}
where $x=0\le r/r_{\rm cl}\le1$ is the normalized impact parameter where $r$ is the projected radius and $r_{\rm cl}$
the cloud radius. The profile of the   mass surface density can then be written as
\begin{equation}
	\label{eq_profile}
	\Sigma_{n} (x) = \frac{2}{\cos^\beta i} r_0\rho_c q^{\frac{n-1}{n}}(1-y_n)^{\frac{1-n}{2}} 
		\int\limits_0^{u_{\rm max}} \frac{{\rm d}u}{(1+u^2)^{n/2}},
\end{equation}
where
\begin{equation}
	u_{\rm max} = \sqrt{\frac{y_n}{1-y_n}}.
\end{equation}
In case of isothermal self-gravitating pressurized clouds the product of inner radius and central density is
proportional to $\sqrt{p_{\rm ext}}$ and is given by
\begin{equation}
	r_0 \rho_c = \sqrt{\frac{\xi_n p_{\rm ext}}{4\pi G}}\frac{1}{\sqrt{q}} ,
\end{equation}
where $p_{\rm ext}$ is the external pressure, $G$ the gravitational constant and where
$\xi_2=2$, $\xi_3=8.63$, and $\xi_4 = 8$ \citep{Fischera2014b}.

\begin{figure}[htbp]
	\includegraphics[width=\hsize]{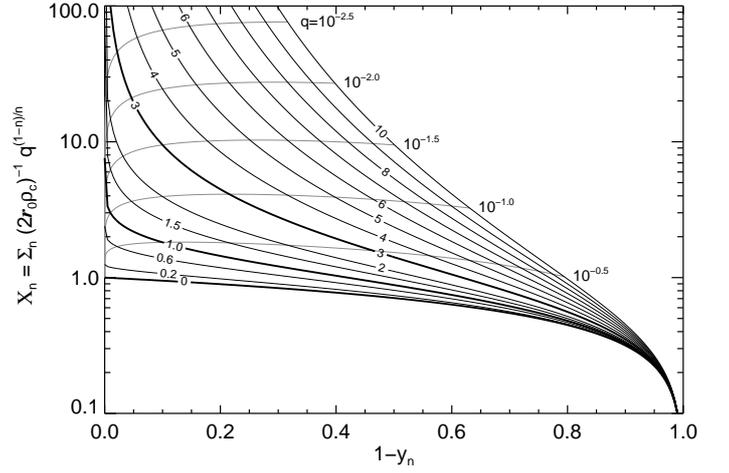}
	\caption{\label{fig_xnyn}Normalized mass surface density $X_n$ as function of $1-y_n$
	for truncated density profiles as given in Eq.~\ref{eq_densityprofile} for various 
	power indices $n$ ranging from 0 to 10. The thicker lines
	correspond to $n=0$, 1, and 3. For a fixed pressure ratio $q$ the mass surface
	densities for a given power index $n$ vary from the central value at $1-y_n=q^{2/n}$ to
	zero at the edge of the cloud where $1-y_n=1$. The central values of the normalized mass surface
	density for various pressure ratios $q$ are shown as thin gray lines. 
	}
\end{figure}

It is convenient to consider the normalized mass surface density
\begin{equation}
	\label{eq_xndef}
	X_n = \Sigma_n \cos^\beta i \,(2\rho_{\rm c}r_0)^{-1}q^{-\frac{n-1}{n}},
\end{equation}
which depends only on the parameter $y_n$. 
The functional dependence is shown in Fig.~\ref{fig_xnyn}.

\begin{figure}[htbp]
	\includegraphics[width=\hsize]{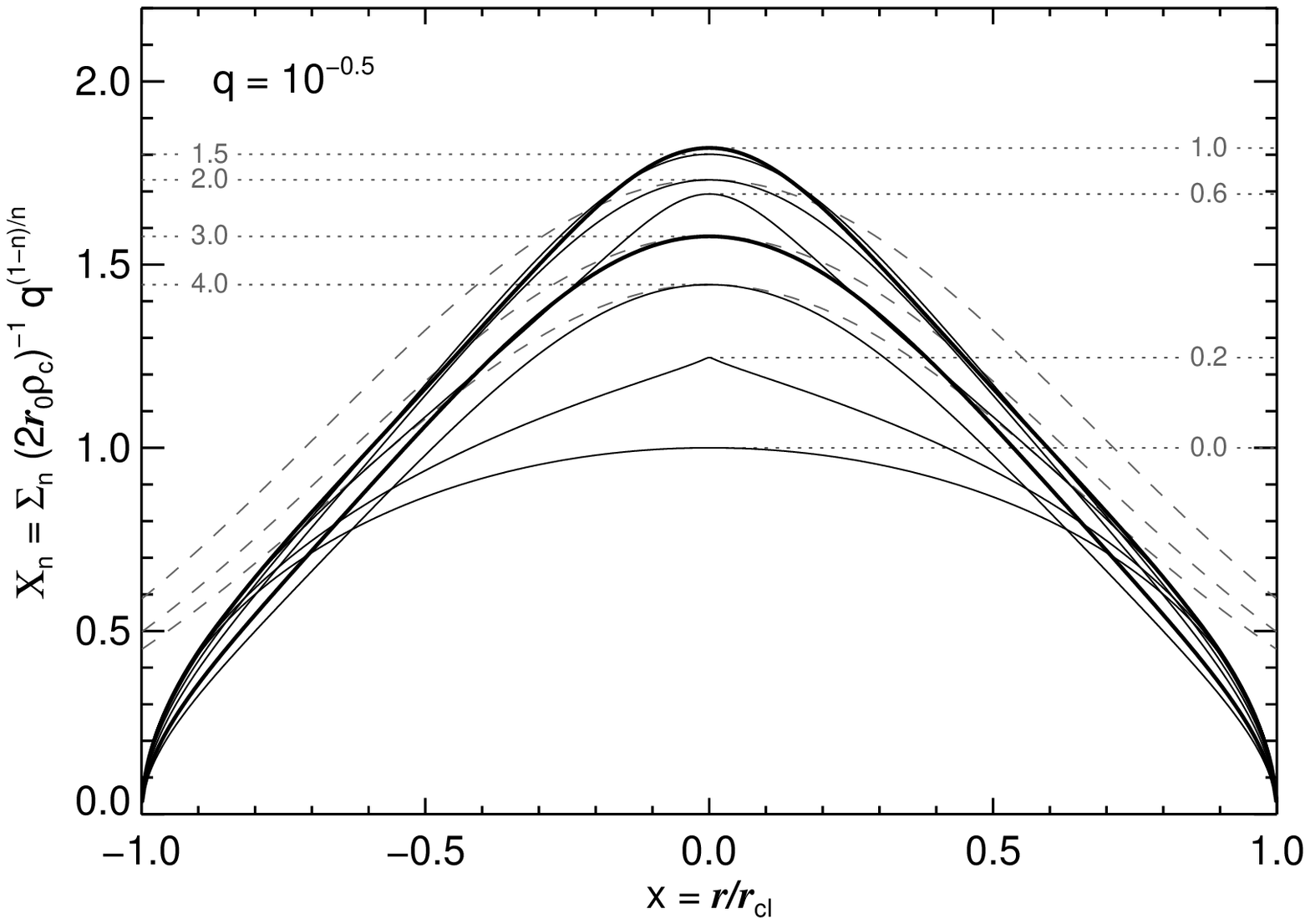}

	\hfill	\includegraphics[width=0.98\hsize]{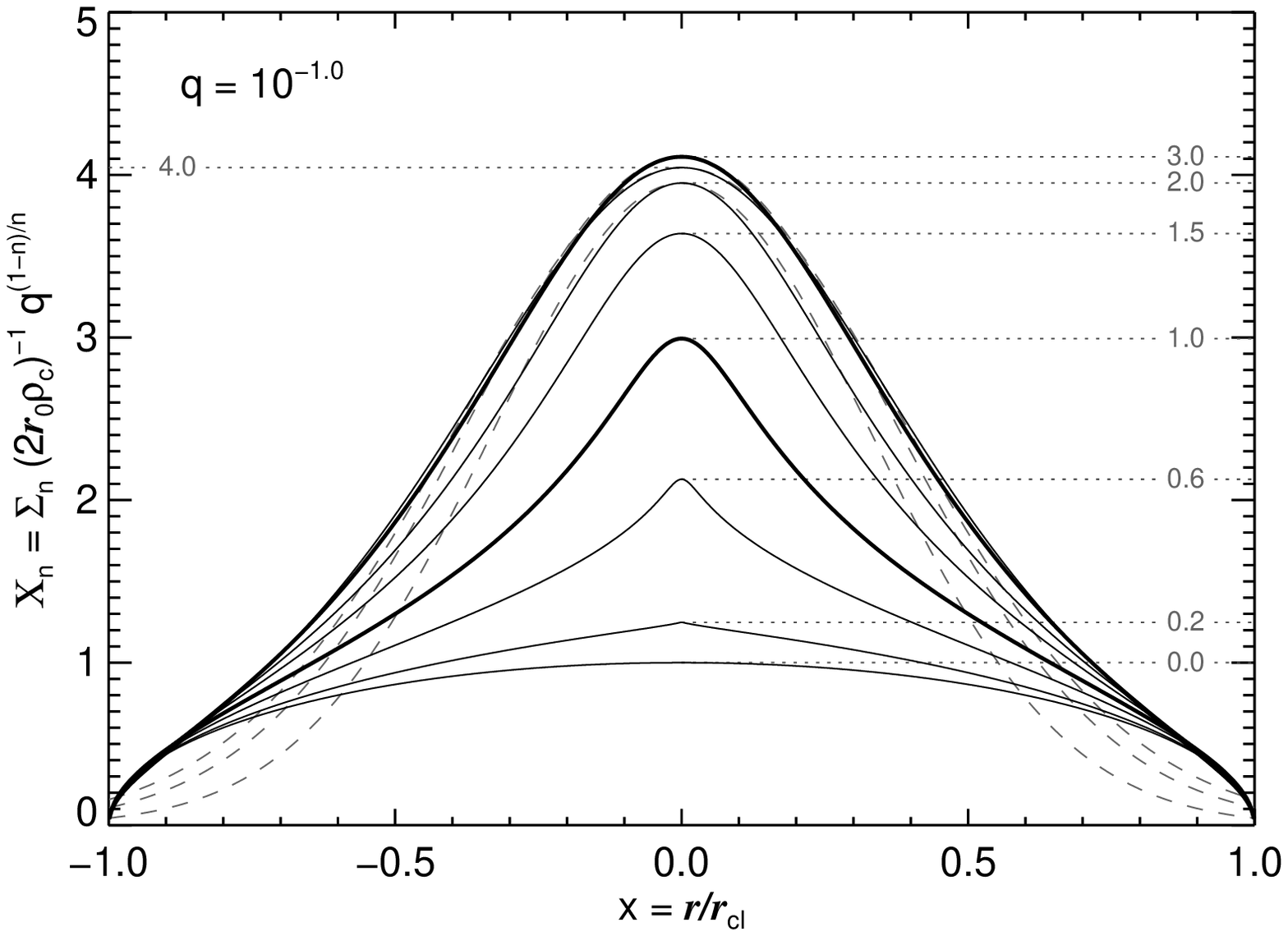}

	\includegraphics[width=\hsize]{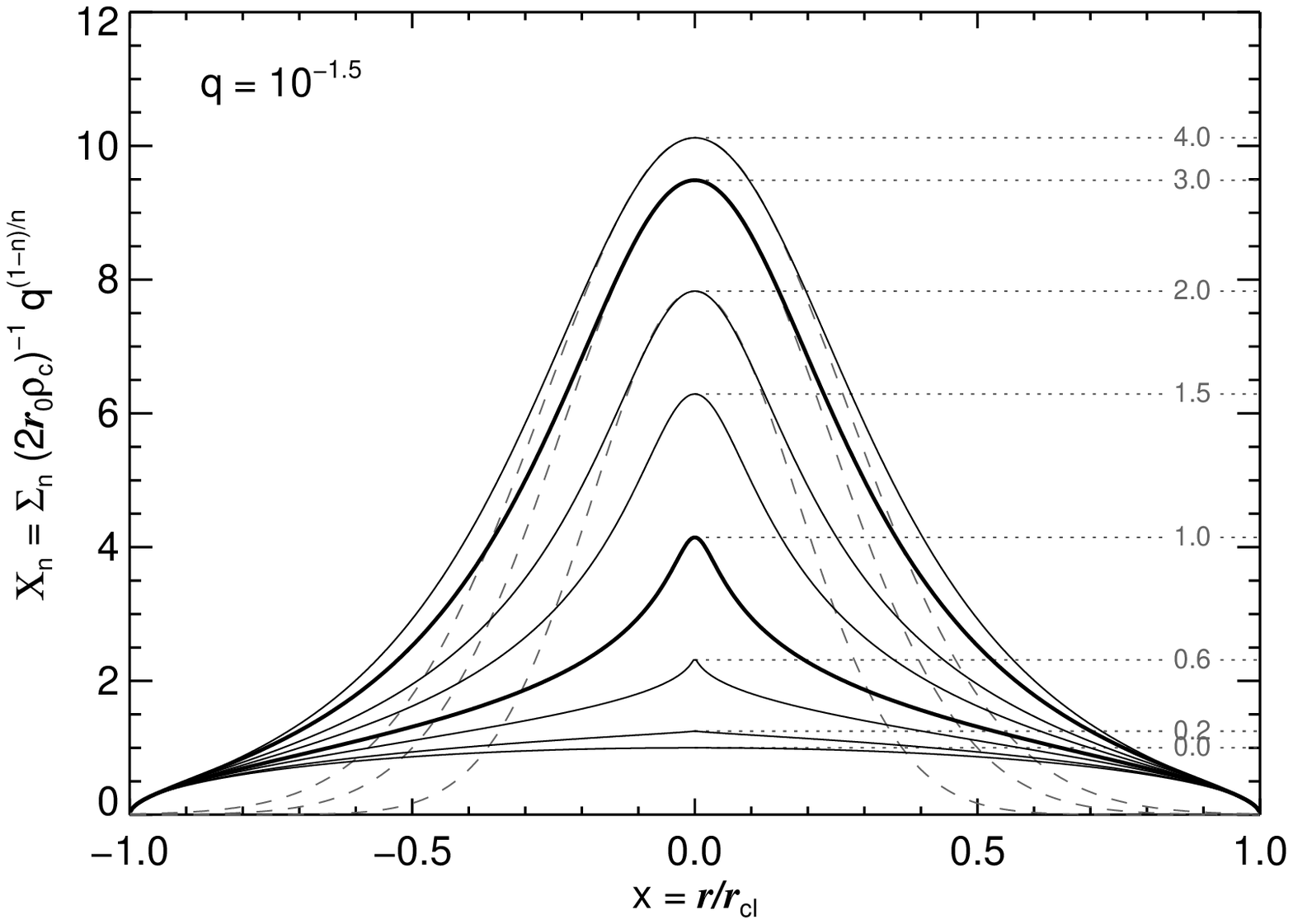}
	\caption{\label{fig_cloudprofile}Profiles of the normalized mass surface density $X_n$
	of a truncated density profile as given in Eq.~\ref{eq_densityprofile} for {three}
	assumptions of the pressure ratio $q=p_{\rm ext}/p_{\rm c}$.
	The maxima or the curves are labeled with the corresponding power index
	$n$. The profiles for $n=1$ and $n=3$ are enhanced with a thicker line.
	The dashed lines are Gaussian approximations (Eq.~\ref{eq_gaussapprox}) where the variance
	is obtained using Eq.~\ref{eq_gaussvariance}. 
	The profile for $n=0$ need to be considered as an asymptote.}
\end{figure}

Profiles of the normalized mass surface density for a number of different assumptions for the power index $n$
of a truncated density profile are shown in Fig.~\ref{fig_cloudprofile}. The method used to derive the profiles
is described in Sects. \ref{sect_msurf_ngt1} and \ref{sect_msurf_nlt1}. The profiles
for $n=1$, 2, 3, and 4 are simple analytical functions given in App.~\ref{app_msurfanalytical}.
At an overdensity of 10 the
inner part of the profiles of cylinders and spheres closely resembles a Gaussian approximation with a width
as given in Sect.~\ref{sect_gaussapprox}. 

\subsection{PDF of the mass surface density}

The PDF of the mass surface density can be given as an implicit function of $y_n$. 
 In the limit $y_n \rightarrow 1$ and $y_n \rightarrow 0$
also explicit expressions of the asymptotic behavior of the PDF can be given and are discussed.

The PDF for the   mass surface density is given by
\begin{equation}
	\label{eq_probability}
	P( \Sigma_{n} (x)) = \frac{{\rm d}P}{{\rm d} \Sigma_{n} }
		= - P(r) \left(\frac{{\rm d} \Sigma_{n} }{{\rm d}r}\right)^{-1},
\end{equation}
where $P(r)$ is the probability to measure a   mass surface density at impact radius $r$.
For a sphere this is given by $P(r) = 2\pi r / (\pi r_{\rm sph}^2)=2x/r_{\rm sph}$ and for a cylinder $P(r) = 1/r_{\rm cyl}$.
{It is convenient to consider the PDF of the normalized mass surface density as defined in Eq.~\ref{eq_xndef} which
is given by
\begin{equation}
	\label{eq_pdfxndef}
	P(X_n) = P(\Sigma_n)\frac{1}{\cos^\beta i}( 2 r_0 \rho_{\rm c})  q^{\frac{n-1}{n}}.
\end{equation}
with $\beta=0$ for spheres and $\beta=1$ for cylinders.}

{By taking the derivative of Eq.~\ref{eq_profile} it is straightforward to show that
\begin{equation}
	\frac{{\rm d} X_n }{{\rm d}r} = -(1-q^{2/n})\frac{x}{r_{\rm cl}} \frac{1+ (n-1)\sqrt{y_n}X_n}{\sqrt{y_n}(1-y_n)}.
\end{equation}
For spheres we obtain the implicit function
\begin{equation}
	\label{eq_pdfsphere}
	P_{\rm sph}(X_n) = \frac{2}{1-q^{2/n}}\frac{\sqrt{y_n}(1-y_n)}{1+(n-1) \sqrt{y_n} X_n(y_n)}.
\end{equation}
As we see the normalized PDF $(1-q^{2/n})P_{\rm sph}(X_n)$ is an implicit function of the 
parameter $y_n$ alone. 
The corresponding PDF for cylinders is 
\begin{equation}
	\label{eq_pdfcylinder1}
	P_{\rm cyl}(X_n) = \frac{1}{2x}P_{\rm sph}(X_n),
\end{equation}
where the normalized impact parameter can be expressed through
\begin{equation}
	\label{eq_impactparameter}
	x = \sqrt{\frac{1-y_n-q^{2/n}}{1-q^{2/n}}}.
\end{equation}
The PDF has a pole at the maximum   mass surface density ($x=0$) or where $y_n = 1-q^{2/n}$ (Fig.~\ref{fig_pdfcyl}). 

}

{Because of the pole for} cylinders we have therefore not a
generalized form of the PDF that depends only on the {parameter} $y_n$. However,  to obtain an
expression which allows a similar discussion in the following section for spheres and cylinders 
we can consider the asymptotic PDF for infinitely high overdensity. 
In the limit $1-y_n\gg q^{2/n}$ the impact parameter behaves as
\begin{equation}
	x \approx \sqrt{1-y_n}.
\end{equation}
We consider therefore the following asymptotic PDF
\begin{equation}
	\label{eq_pdfcylinder}
	P^{(a)}_{\rm cyl}(X_n) = \frac{1}{1-q^{2/n}}\frac{\sqrt{y_n}\sqrt{1-y_n}}{1+(n-1) \sqrt{y_n} X_n(y_n)}.
\end{equation}
As shown in App.~\ref{app_pdfasymptotes} 
the asymptotic PDF provides for all $q$ the correct asymptote in the limit of small mass surface densities
 and the correct asymptotic behavior at large mass surface densities for $q\rightarrow 0$.
  
\section{\label{sect_properties}Characteristics of the PDF}

\begin{figure}
	\includegraphics[width=\hsize]{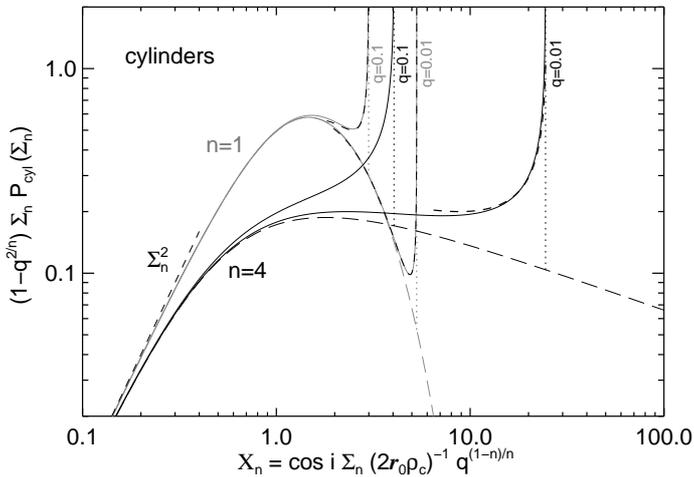}
	\caption{\label{fig_pdfcyl} PDF of the normalized   mass surface density $X_n$ of cylindrical clouds having
	a density distribution with $n=1$ (gray lines) and $n=4$ (black lines). The clouds have an overdensity ($1/q$)
	of either 10 or 100. The long dashed curves show the asymptote for infinite high overdensity. The dotted lines mark
	the asymptotic value at the poles. The short dashed lines give the asymptotic values in the limit of high and low
	values of $X_n$.}
\end{figure}

\begin{figure*}[htbp]
	\hfill
	\includegraphics[width=0.9\hsize]{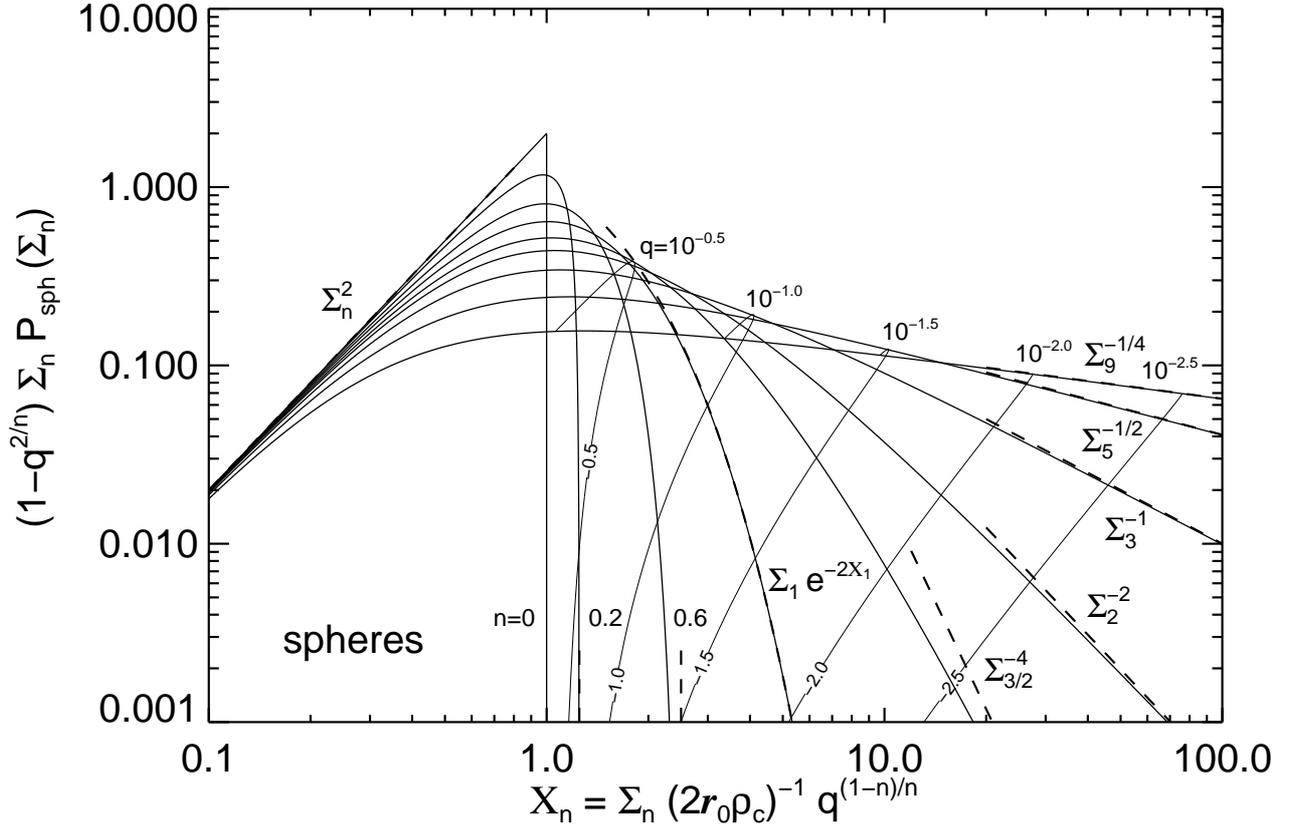}
	\hfill
	\hfill\\[0.2cm]
	
	\hfill
	\includegraphics[width=0.9\hsize]{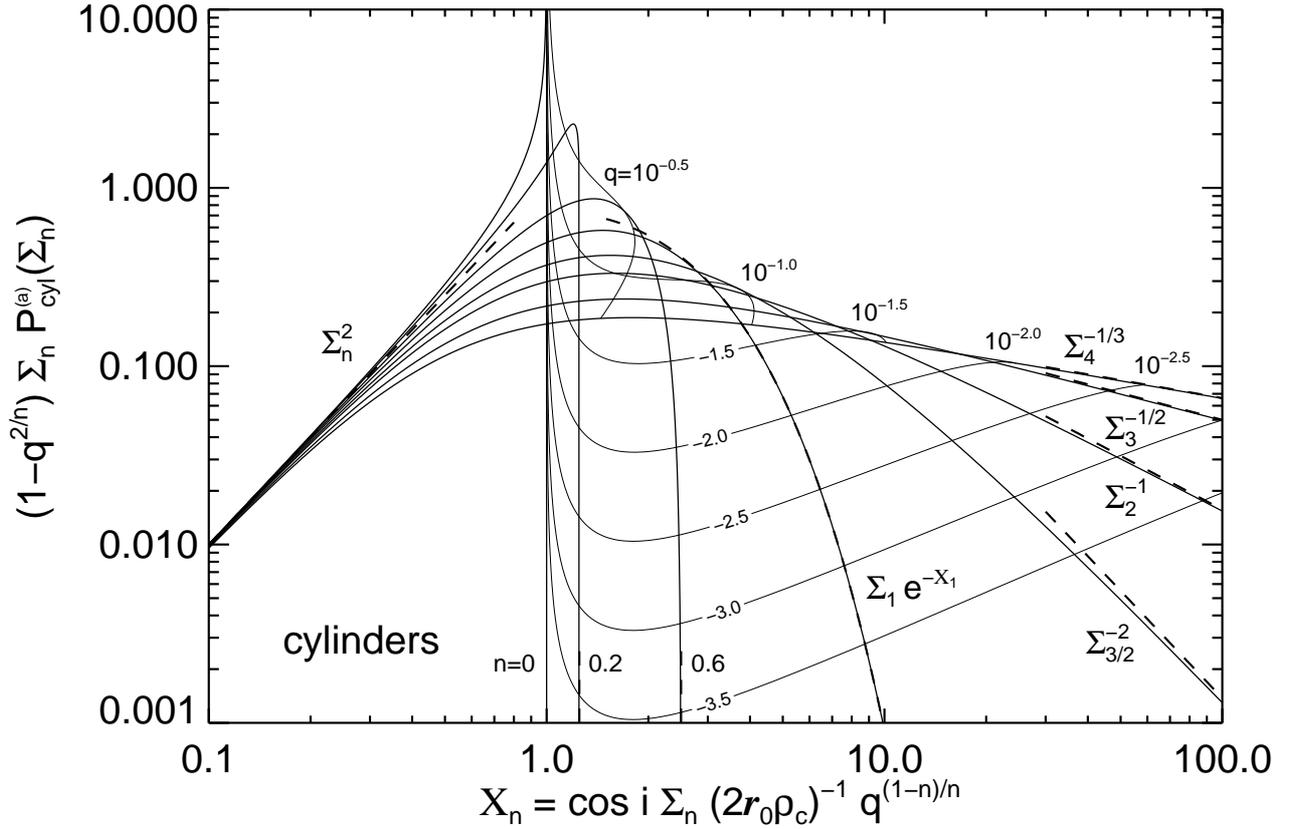}
	\hfill
	\hfill
	\caption{\label{fig_pdf}
	PDF of the normalized   mass surface density $X_n$ in case of spheres (top figure) 
	and the asymptotic PDF in case of cylinders (bottom figure) with a truncated radial density profile 
	given by Eq.~\ref{eq_densityprofile} for various assumptions of the power $n$. 
	The thin lines show
	at which   mass surface density $X_n$ at given density ratio $q=\rho(r_{\rm cl})/\rho_{\rm c}$ the PDF corresponding
	to a certain power $n$ truncates (for spheres) or has a pole (for cylinders). 
	The lines are labeled with $\log_{10} q$. The dashed lines show
	the asymptotic behavior at large and small $X_n$.
	}
\end{figure*}

{The PDFs of spheres and the asymptotic PDFs of cylinders with a truncated analytical 
density profile for various different assumptions for the power $n$ and the pressure ratio $q$ 
are shown in Fig.~\ref{fig_pdf}. They are truncated at the highest mass surface 
density. 

The exact asymptotes for high and low mass surface densities also
shown in the figure are derived in Sect.~\ref{app_pdfasymptotes}. The maximum position is discussed
in more detail in Sect.~\ref{app_pdfmaxima}.

The curves have a functional form that depends only on the power $n$ of the radial density profile.
They are truncated at the central mass surface density.

\subsection{Asymptotes at high and low mass surface densities}
At low mass surface densities the PDF  approaches asymptotically a power law
$P(\Sigma)\propto \Sigma$. The asymptotic behavior at high mass surface densities depends on
the steepness of the radial density profile. 
For $n>1$ the PDF approaches asymptotically a power
law given by 
\begin{equation}
	P_{\rm sph}(X_n) \propto X_n^{-\frac{n+1}{n-1}},\quad P_{\rm cyl}^{(a)}(X_n)\propto X_n^{-\frac{n}{n-1}}.
\end{equation}
As can be seen in the figure the asymptote is a better representation of the PDF at high mass surface densities for
steeper density profiles.
In the limit of large power $n$ the PDF at high mass surface densities becomes 
$P(X_n)\propto X_n^{-1}$ as expected for a source with a Gaussian density profile (Sect.~\ref{sect_pdfgauss}).
For $n=1$ the PDF at high mass surface density is an exponential function 
\begin{equation}
	P_{\rm sph}(X_n)\propto e^{-2 X_1}, \quad P_{\rm cyl}^{(a)}(X_n) \propto e^{-X_1}. 
\end{equation}
For clouds with $n<1$ the PDF is limited to a maximum mass surface
density given by $X_n=1/(1-n)$. In the neighborhood of this maximum 
value the PDF varies strongly with mass surface density. 
In the limit of $n=0$ the PDF becomes identical to the PDF of a homogeneous sphere or cylinder. 

The power law asymptote at high mass surface densities may be used to infer the power $n$ of the 
radial density profile. For spheres a power law $P_{\rm sph}\propto \Sigma^{-\alpha}$ would indicate a power 
index $n=(\alpha+1)/(\alpha-1)$ of the radial density profile as can be derived for simple power law density profiles
\citep{Kritsuk2011,Federrath2013a}.
As shown in Fig.~\ref{fig_pdfcyl} the asymptotic behavior at large mass surface densities 
in case of cylinders is only established for sufficiently high overdensities. For example, a profile with $n=4$ which
is consistent with the density profile of self-gravitating isothermal cylinders the PDF for overdensities as
high as 100 has no apparent power law at high mass surface densities. For cylinders with high overpressure
the PDF at the pole is approximately given by $P_{\rm cyl}(X_n)\propto X_n^{-\frac{n}{n-1}}/\sqrt{1-(X_n/X_n(0))^{2/(n-1)}}$
A power law is only established for the range $[X_n]_{\rm max}\ll X_n\ll X_n(0)$ where $[X_n]_{\rm max}$ is
the mass surface density at the PDF maximum (Sect.~\ref{sect_pdfmaxima}).
}

\subsection{\label{sect_pdfmaxima}The maxima of the asymptotic PDF}

{As we see in Fig.~\ref{fig_pdf} 
the PDF of spheres and the asymptotic PDF of cylinders with the analytical density profile
have well defined maxima at $[X_n]_{\rm max}$ which depend only on the power $n$.
This allows a simple interpretation of the observed mass surface density at the peak position
in terms of the normalization factor $2r_0\rho_{\rm c}q^{(n-1)/n}$ for given $n$
using the definition Eq.~\ref{eq_xndef}. 
In case of isothermal clouds the maximum position can 
be used to infer the pressure in the ambient medium.}

\begin{table}
	\caption{\label{table_pdfmaxima}Maxima positions of the linear and logarithmic PDF}
	\begin{tabular}{cccccc}
		& & \multicolumn{2}{c}{sphere $P_{\rm sph}(X_n)$} & \multicolumn{2}{c}{cylinder $P^{(a)}_{\rm cyl}(X_n)$} \\
		n & PDF & $[y_n]_{\rm max}$ & $[X_n]_{\rm max}$ & $[y_n]_{\rm max}$ & $[X_n]_{\rm max}$ \\
		\hline
		1 & $P(X_n)$ & $\frac{1}{3}$ & $\ln \frac{\sqrt{3}+1}{\sqrt{2}}$ & $\frac{1}{2}$ & $\ln [1+\sqrt{2}]$ \\
		2 & $P(X_n)$ & 0.2110 & 0.5373 & 0.2723 & 0.7277 \\
		3 & $P(X_n)$ & $\frac{2-\sqrt{3}}{\sqrt{3}}$ & $\frac{\sqrt{2\sqrt{3}-3}}{2\sqrt{3}-2}$ 
				& $\frac{\sqrt{33}-5}{4}$ & $2\frac{\sqrt{\sqrt{33}-5}}{9-\sqrt{33}}$ \\
		4 & $P(X_n)$ & 0.1222 & 0.4162 & 0.1413 & 0.4609 \\
		\hline
		1 & $X_n P(X_n)$ & 0.5861 & 1.0096 & 0.8069 & 1.4633 \\
		2 & $X_n P(X_n)$ & 0.4780 & 1.0566 & 0.6547 & 1.6041 \\
		3 & $X_n P(X_n)$ & $\sqrt{2}-1$ & $\frac{\sqrt{2\sqrt{2}-2}}{2\sqrt{2}-2}$ & $\frac{\sqrt{17}-3}{2}$ 
				& $\frac{\sqrt{2\sqrt{17}-6}}{5-\sqrt{17}}$ \\
		4 & $X_n P(X_n)$ & 0.3700 & 1.1367 & 0.4970 & 1.7974 
	\end{tabular}
\end{table}

\begin{figure*}[htbp]
	\hfill
	\includegraphics[width=0.46\hsize]{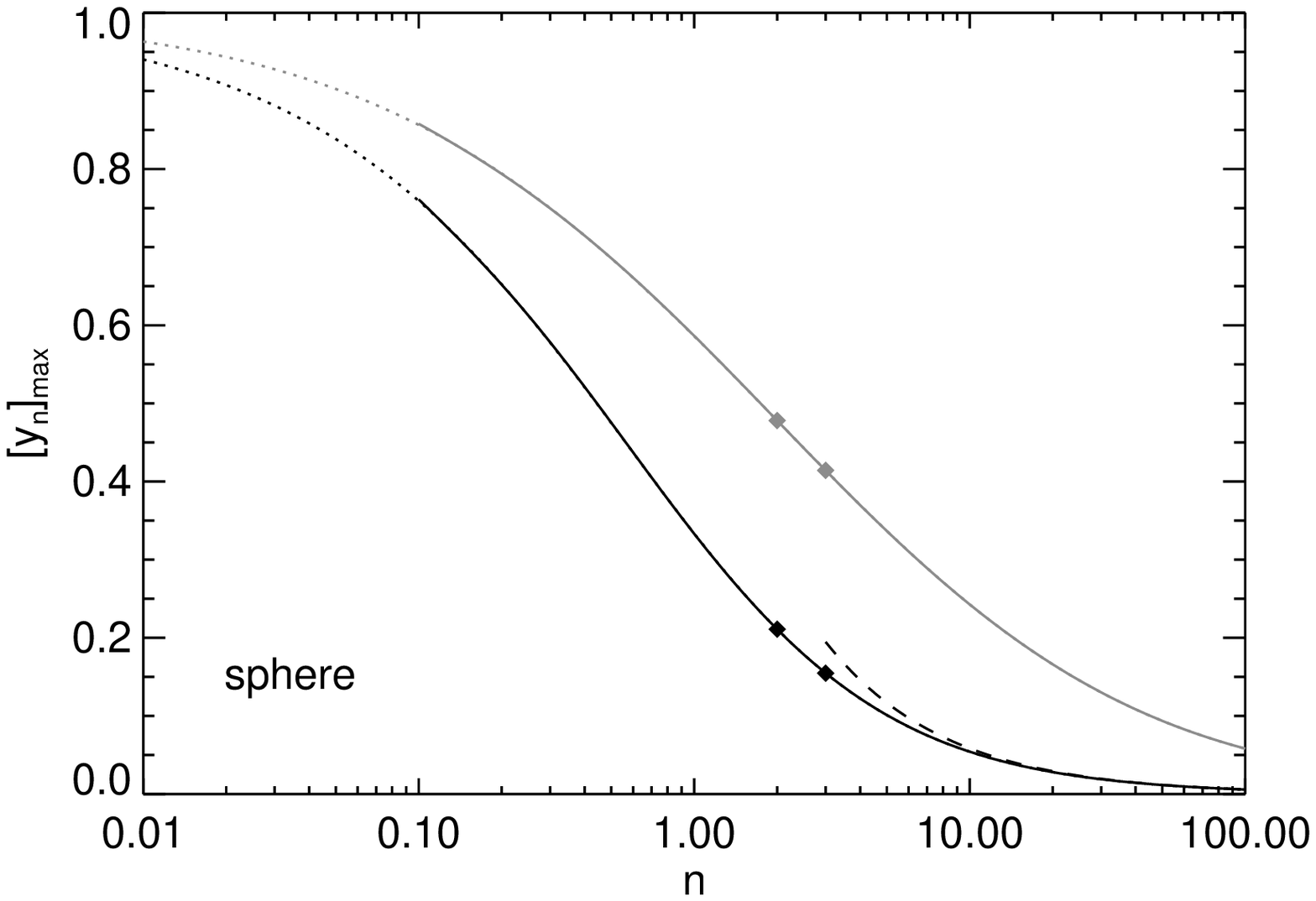}
	\hfill
	\includegraphics[width=0.46\hsize]{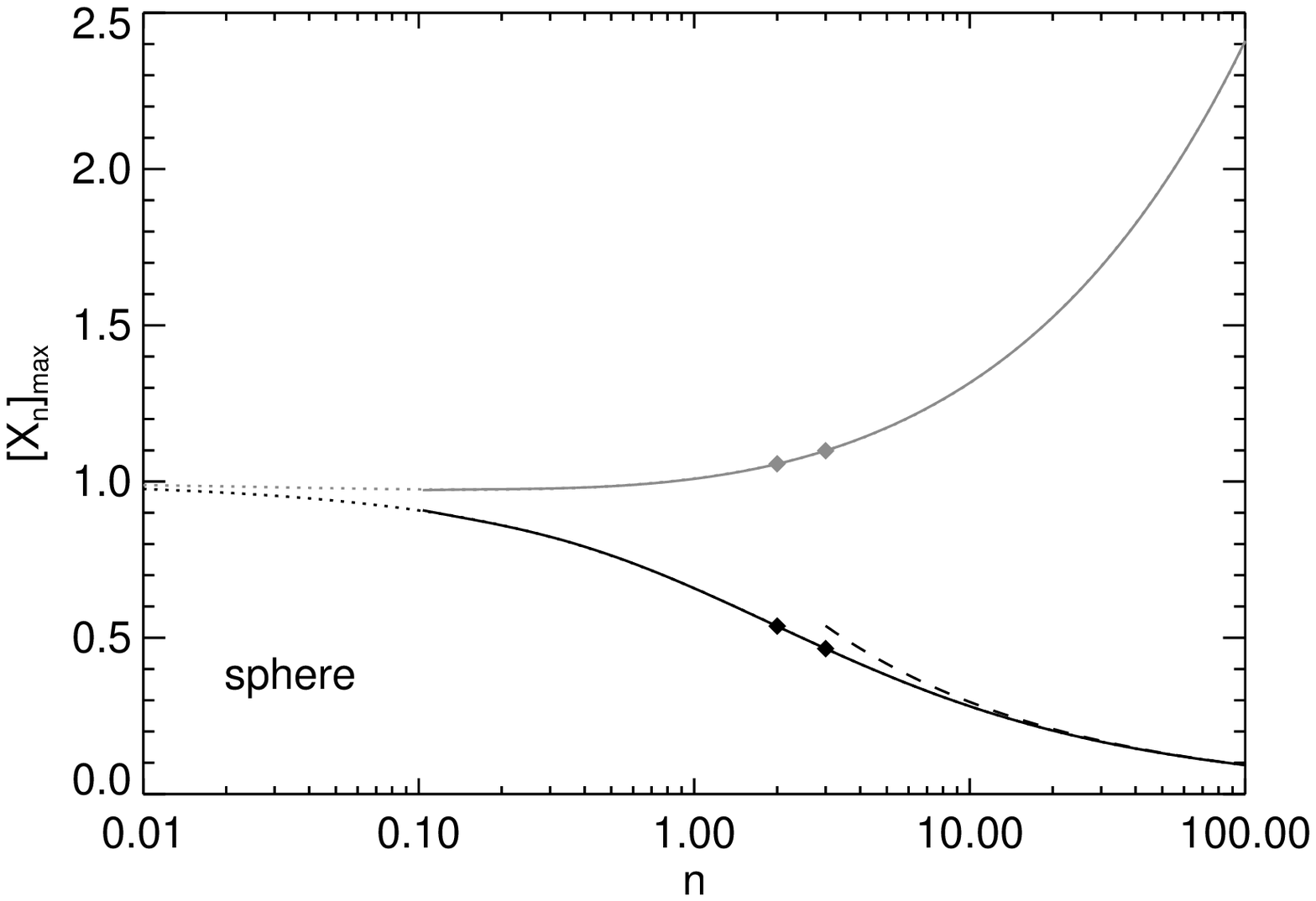}\hfill\hfill\\[0.1cm]

	\hfill
	\includegraphics[width=0.46\hsize]{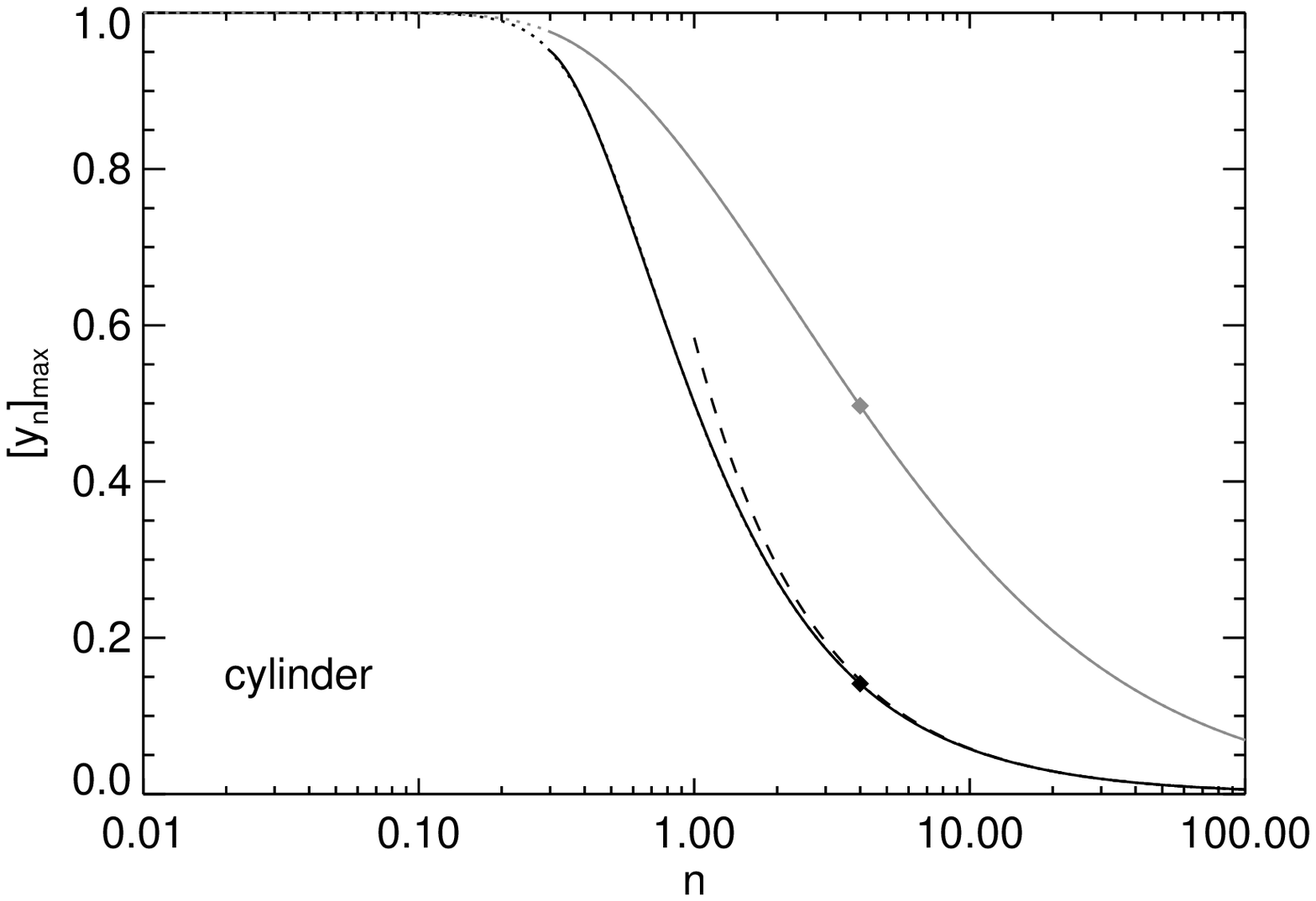}
	\hfill
	\includegraphics[width=0.46\hsize]{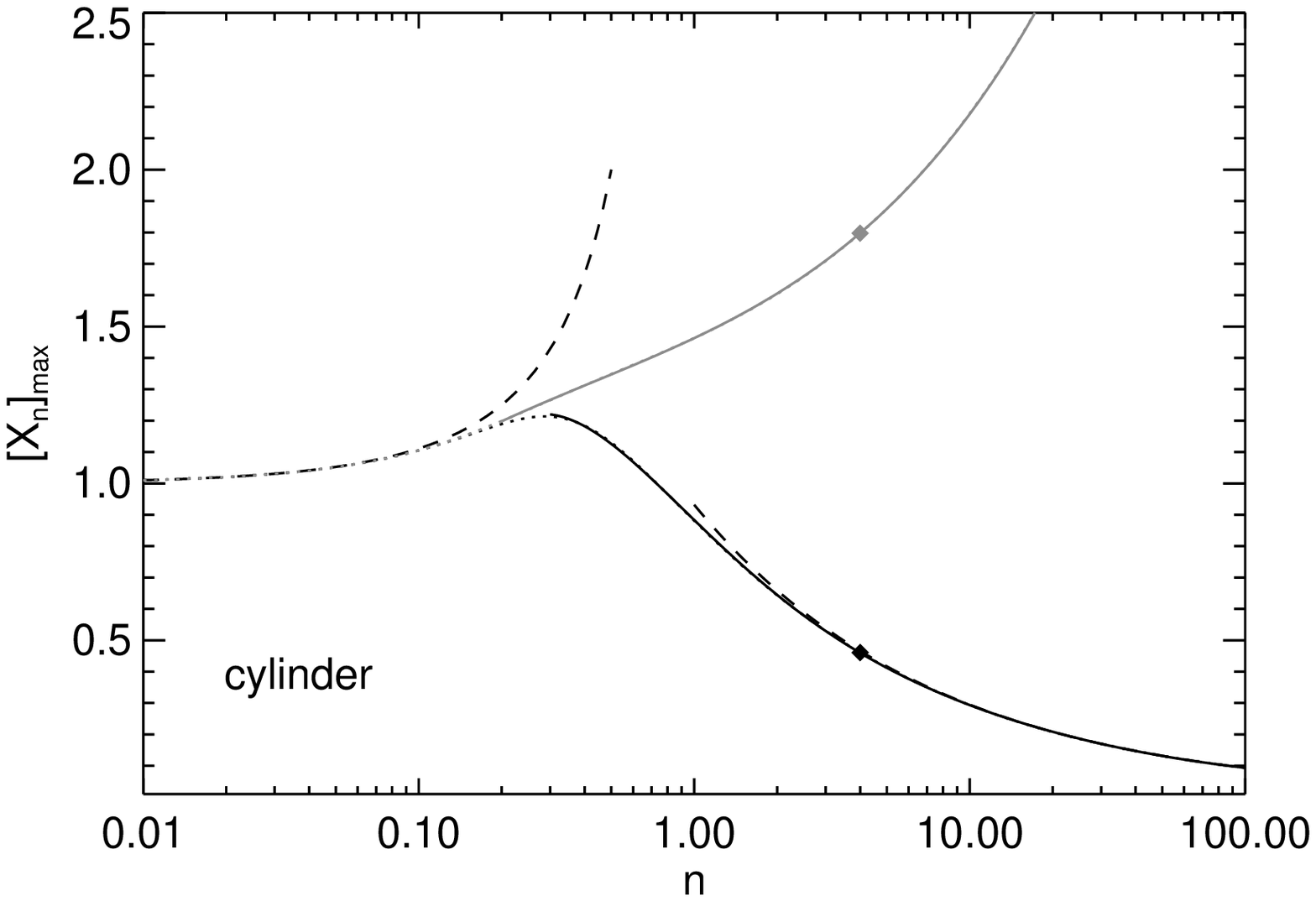}\hfill\hfill
	\caption{\label{fig_pdfmaxima}Maxima position of the PDF of the parameter $y_n$ and the
	normalized   mass surface density $X_n$ for spheres (top) and cylinders (bottom) as function
	of the power $n$ (dotted curves). Black lines correspond to the linear PDF and the gray lines
	to logarithmic PDF. The solid curves are polynomial fits to the accurate position values. The
	values for isothermal self-gravitating spheres ($n=2$ and $n=3$) and cylinders ($n=4$) 
	are shown as small diamonds. The asymptotic behavior
	of $[X_n]_{\rm max}=1/(1-n)$ for $n\rightarrow 0$ for cylinders is shown as dashed line. 
	{Asymptotes described in Sect.~\ref{app_maxasymptote} are shown as dashed lines.}}
\end{figure*}


{Related to the maximum position is a parameter $[y_n]_{\rm max}$.
The functional dependence of the maximum position $[y_n]_{\rm max}$ and $[X_n]_{\rm max}$ 
of the linear and logarithmic asymptotic PDF on the power $n$
is shown in Fig.~\ref{fig_pdfmaxima}. The curves are derived by solving the conditional equations
given in App.~\ref{app_pdfmaxima}. Selected values are given in Tab.~\ref{table_pdfmaxima}.}

As shown in Sect.~\ref{sect_approx_n0}, in the limit $n\rightarrow 0$ the PDFs become
the ones of homogeneous spheres and cylinders where the maximum value is related to
the central mass surface density. In this limit 
we have therefore $[y_n]_{\rm max}\rightarrow 1$ {and $[X_n ]_{\rm max}=1$. In case
of cylinders the normalized mass surface density at PDF maximum in the limit of flat density
profiles is approximately given by $[X_n]_{\rm max}=1/(1-n)$.

The parameter $[y_n]_{\rm max}$ provides for given power $n$ an estimate of the corresponding 
impact parameter as function of the pressure ratio $q$ based on Eq.~\ref{eq_impactparameter}. 
Below a minimum overpressure
$(1-[y_n]_{\rm max})^{-n/2}$ the maximum coincides with the central mass surface density.
At higher overpressures the impact parameter moves outwards and reaches asymptotically
a maximum value $x = \sqrt{1-[y_n]_{\rm max}}$.
If we consider for example the logarithmic PDF of a sphere with $n=3$ the minimum overdensity with
$x=0$ is 
\begin{equation}
	q^{-1} = \left(1-\left[\sqrt{2}-1\right]\right)^{-3/2} \approx 2.23,
\end{equation}
where the parameter $[y_n]_{\rm max}$ is taken from Tab.~\ref{table_pdfmaxima}.
In the limit $q\rightarrow 0$ the impact parameter at the PDF maximum becomes
\begin{equation}
	x = \sqrt{1-\left[\sqrt{2}-1\right]} \approx 0.765.
\end{equation}

As the maximum position $[y_n]_{\rm max}$ decreases with $n$ the impact parameter related to
the PDF maximum is larger for steeper density profiles and reaches $x=1$ for $n\rightarrow \infty$.
In case of the logarithmic PDF the mass surface density related to the PDF maximum  
is larger in respect to the linear PDF and corresponds to a smaller impact parameter.



{The} dependence of the maxima of the linear PDF in the limit $n\rightarrow \infty$ can be 
described by simple power laws as shown in App.~\ref{app_maxasymptote}.
The asymptotes for $[y_n]_{\rm max}$ and $[X_n]_{\rm max}$ provide as shown in Fig.~\ref{fig_pdfmaxima}
good results for spheres above $n\approx 3$ and for cylinders above $n\approx 1$.
}

\begin{table*}[htbp]
\center
	\caption{\label{table_pdfmaximafit} Coefficients of the polynomial approximation of the maximum position of the asymptotic PDF}
	\begin{tabular}{ccccccccc}
		&$a_1$& $a_2$& $a_3$ & $a_4$ & $a_5$ & $a_6$ &$n_{\rm min}$ & $n_{\rm max}$\\
		\multicolumn{9}{c}{Spheres, linear values } \\
		$[y_n]_{\rm max}$ & -1.099 & -0.5883 & -0.1071 & 0.004447 & 0.002598 & -0.0003259  & 0.1 & 100.0\\
		$[X_n]_{\rm max}$ & -0.4181 & -0.2544 & -0.05889 & 0.0008424 & 0.001723 & -0.0001919 & 0.1 & 100.0\\
		\multicolumn{9}{c}{Spheres, log values } \\
		$[y_n]_{\rm max}$ & -0.5338 & -0.2590 & -0.04989 & -0.002858 & 0.0004857  & ---& 0.1 & 100.0\\
		$[X_n]_{\rm max}$ & 0.008622 & 0.04882 & 0.02353 & 0.003190 & -0.0003657 & --- & 0.1 & 100.0\\
		\multicolumn{9}{c}{Cylinders, linear values } \\
		$[y_n]_{\rm max}$ & -0.6920 & -0.8029 & -0.1382 & 0.05612 & -0.01153 & 0.0009113 & 0.3 & 100.0\\
		$[X_n]_{\rm max}$ & -0.1243 & -0.4190 & -0.07131 & 0.03418 & -0.007760 & 0.0006513 & 0.3 & 100.0\\
		\multicolumn{9}{c}{Cylinders, log values } \\
		$[y_n]_{\rm max}$ & -0.2146 & -0.2506 & -0.07477 & 0.002226 & 0.0001414 & --- & 0.3 & 100.0\\
		$[X_n]_{\rm max}$ & 0.3806 & 0.1235 & 0.01110 & 0.006071 & -0.0007028 & --- & 0.2 & 100.0\\
	\end{tabular}
\center
\end{table*}

The ratio
\begin{equation}
	\frac{[\Sigma_n]_{\rm cut}}{[\Sigma_n]_{\rm max}}=\frac{[X_n]_{\rm cut}}{[X_n]_{\rm max}} 
\end{equation}
between the central   mass surface density $[X_n]_{\rm cut}$ and the   mass surface at the PDF maximum
can be used to infer the pressure ratio $q$ for given power $n$.
{As an example we consider a critical stable sphere which radial density profile is close to the analytical 
profile with $n=3$. The critical stable sphere has a density ratio of $q\approx 1/14.04$
Applying Eq.~\ref{eq_msurfisosphere} we find that the corresponding central mass surface density is given by
\begin{equation}
	[X_3]_{\rm cut} =  q_{\rm crit}^{-\frac{2}{3}}\sqrt{1-q_{\rm crit}^{2/3}} \approx 5.296.
\end{equation}
For critical stable spheres the ratio between the cutoff and the maximum of the logarithmic PDF is therefore given by
\begin{equation}
	\frac{[X_3]_{\rm cut}}{[X_3]_{\rm max}} = q_{\rm crit}^{-\frac{2}{3}}\sqrt{1-q_{\rm crit}^{2/3}}\frac{2\sqrt{2}-2}{\sqrt{2\sqrt{2}-2}}\approx 4.821,
\end{equation}
where the maximum position is taken from Tab.~\ref{table_pdfmaxima}.
}

The functional dependence of the maximum position on the power $n$ can over a large range
be well fitted by a polynomial function 
\begin{eqnarray}
	[y_n]_{\rm max} &=& \sum_{i=1}^{N_{\rm max}} a_i (\ln n)^i, \\\
	[X_n]_{\rm max} &=& \sum_{i=1}^{N_{\rm max}} a_i (\ln n)^i,
\end{eqnarray}
where $N_{\rm max}=6$ for spheres and $N_{\rm max}=5$ for cylinders. 
The coefficients are listed in Tab.~\ref{table_pdfmaximafit}.

\section{Summary and conclusion}

The study summarizes a series of properties of the PDF of the  mass surface
density of spherical and cylindrical structures {having an analytical radial density
profile $\rho =\rho_{\rm c}/(1+(r/r_0)^2)^{n/2}$ where $\rho_c$ is the central density
and $r_0$ the inner radius.}
The profiles are assumed to be truncated at a cloud radius $r_{\rm cl}$ 
as expected for cold structures
embedded in a considerably warmer medium. 
The results are therefore applicable to
individual condensed structures in star forming molecular clouds. 

{The PDF for given geometry is determined by the power $n$, the
density ratio $q=\rho(r_{\rm cl})/\rho_{\rm c}$, the product $\rho_{\rm c}r_{0}$,
and, in case of a cylinder, the inclination angle $i$.
It is convenient to describe the properties of the PDF in terms of the 
unit free mass surface density defined by
\begin{equation}
	X_n = \Sigma_n \frac{\cos^\beta i}{2\rho_{\rm c} r_0}q^{-\frac{n-1}{n}},
\end{equation}
where $\beta=0$ for spheres and $\beta=1$ for cylinders. 
The properties are:}
\begin{enumerate}
	\item {For given geometry and power index $n$
	the normalized PDF $(1-q^{2/n})P(X_n)$ is a unique curve expressed
	through a simple implicit function of the parameter $y_n = (1-q^{2/n})(1-x^2)$
	where $x = r/r_{\rm cl}$ is the normalized impact parameter.}
	\item At the central  mass surface density $X_n(0)$
	the PDF of spheres has a sharp cut-off while the PDF of cylinders has a pole.
	\item {At high overdensities the PDF  
	has a well defined maximum at fixed $[X_n]_{\rm max}$.} 
	\item At  mass surface densities which are small relative to the maximum {position}
	the asymptotic PDF approaches asymptotically a power law
	$P(X_{\rm n})\propto X_n$.
	\item {In the limit of high overdensities the PDFs 
	approach for $n>1$ at mass surface densities above the
	peak} asymptotically power laws. They
	are given by $P(X_n)\propto X_n^{-\frac{n+1}{n-1}}$  {in case of spheres and, with the exception 
	of the pole,
	by $P(X_n)\propto X^{-\frac{n}{n-1}}$ in case of cylinders. For given overdensity the asymptote is
	a better approximation for steeper density profiles (larger $n$).}
	\item {For $n<1$ the PDF has a strong cutoff and is limited to a maximum  mass surface density 
	$X_n\le 1/(1-n)$.}
\end{enumerate}

The slope {of the PDF} at high mass surface densities 
can also be obtained assuming a simple power law profile for spherical
clouds {(e.g. \citealt{Kritsuk2011,Federrath2013a}).} 
But it should be considered that this profile only is an asymptotic behavior {in}
the limit of high overdensities
and seems more appropriate for collapsing clouds while most condensations might not
be in such a state. As shown in the paper in general the {shape of the PDF} is not a power law. 
Further, the profile would produce a nonphysical high probability at low 
mass surface densities. 



The derived properties are related to background subtracted structures within molecular clouds.
They are therefore not directly applicable to measurements of the global PDF of molecular
clouds which is a statistical mean of different properties not only of the condensed structures but
the surrounding medium as well. For instance the tail at high  mass surface densities 
seen in the PDF of star forming molecular clouds may have different physical explanations. 
It also need to be considered that the functional form
of the PDF is affected by an additional background. In case of filaments the situation is furthermore 
complicated through a possible variation of the inclination angle. 
Those problems are addressed in a following paper \citep{Fischera2014b} based on isothermal
{self-gravitating} pressurized spheres and cylinders. 

\begin{acknowledgement}
	This work was supported by grants from the Natural
  	Sciences and Engineering Research Council of Canada and the Canadian
  	Space Agency. The author likes to thank Prof. P. G. Martin and Dr. Richard Tuffs 
	for his support, Quang Nguyen Luong for reading the manuscript and his 
	helpful comments, and the unknown referee for the suggestions.
\end{acknowledgement}

\appendix

\section{Solution for the mass surface density profile}

\subsection{\label{sect_msurf_ngt1}Case $n\ge 1$}

In case $n>1$ the integral can be expressed through the incomplete and complete beta function
\begin{eqnarray}
	\label{eq_integralbeta}
	\int\limits_0^{u_{\rm max}} \frac{{\rm d}u}{(1+u^2)^{\frac{n}{2}}} &=& \frac{1}{2}{\rm B}\left(\frac{n}{2}-\frac{1}{2},\frac{1}{2}\right)\nonumber\\
		&&\times\bigg\{1-I_{1-y_n}\left(\frac{n}{2}-\frac{1}{2},\frac{1}{2}\right)\bigg\},
\end{eqnarray}
with the condition $a,b>0$ where the normalized incomplete beta function is given by
\begin{equation}
	I_\xi(a,b) = \frac{1}{B(a,b)}\int_0^\xi{\rm d}t\,t^{a-1}(1-t)^{b-1}.
\end{equation}
The beta function is equal to
\begin{equation}
	{\rm B} (a,b) = \frac{\Gamma(a)\Gamma(b)}{\Gamma(a+b)},
\end{equation}
where $\Gamma(x)$ is the $\Gamma$-function given by
\begin{equation}
	\Gamma(x) = \int_0^{\infty}{\rm d}t\,t^{x-1} e^{-t}.
\end{equation}
In Sect.~\ref{sect_incompletebeta} the asymptotic behavior of $I_\xi(a,b)$ with $a=(n-1)/2$ and $b=0.5$ is discussed.
In general the asymptotic behavior is better for higher powers of $a$.

The   mass surface density for given external pressure and overdensity through the cloud center for $n>1$ is
\begin{equation}
	\Sigma_n(0) = \frac{r_0\rho_c}{\cos^\beta i} \,
		{\rm B}\left(\frac{n-1}{2},\frac{1}{2}\right)\left\{1-I_{q^{2/n}}\left(\frac{n-1}{2},\frac{1}{2}\right)\right\}.
\end{equation}
In the limit of high overdensity ($q\rightarrow 0$) the central   mass surface density becomes the asymptotic value
\begin{equation}
	\label{eq_msurf0approx}
	 \Sigma_n (0) \approx \frac{r_0\rho_c}{\cos^\beta i}  \frac{\Gamma(\frac{n-1}{2})\Gamma(\frac{1}{2})}{\Gamma(\frac{n}{2})}.
\end{equation}





\subsection{\label{sect_msurf_nlt1}Case $n<1$}

To estimate the profile for $n<1$ we can transform the integral to
\begin{eqnarray}
	\label{eq_integraltransform}
	\int\limits_0^{u_{\rm max}}\frac{{\rm d}u}{(1+u^2)^{\frac{n}{2}}} &=& \frac{1}{n-1} \Bigg\{n \int\limits_0^{u_{\rm max}}\frac{{\rm d}u}{(1+u^2)^{\frac{n}{2}+1}}\nonumber\\
		&&- \frac{u_{\rm max}}{(1+u_{\rm max}^2)^{\frac{n}{2}}}\Bigg\}.
\end{eqnarray}
The integral can then be calculated using the complete and incomplete beta function as in Eq.~\ref{eq_integralbeta}. 
The   mass surface density profile becomes
\begin{eqnarray}
	\label{eq_profile2}
	 \Sigma_{n} (x) &=& \frac{2}{\cos^\beta i} r_0\rho_c q^{\frac{n-1}{n}}\frac{1}{1-n}\Bigg\{ \sqrt{y_n}\nonumber\\
		&&
		-n  (1-y_n)^{\frac{1-n}{2}} \int\limits_0^{u_{\rm max}} \frac{{\rm d}u}{(1+u^2)^{\frac{n}{2}+1}}\Bigg\}.
\end{eqnarray}

In the limit of high overdensity the central mass surface density approaches asymptotically a maximum value given by
\begin{equation}
	 \Sigma_n (0) = \frac{2}{\cos^\beta i} r_0\rho_c q^{\frac{n-1}{n}}\frac{1}{1-n}.
\end{equation}

\subsection{\label{app_msurfanalytical}Analytical profiles of the mass surface densities}

The mass surface density profiles of the truncated analytical density profile 
for any natural number $n=1,2,...$ can be expressed through simple analytical
functions. For $n=1$, $2$, $3$, and $4$ 
the profiles are for example given by
\begin{eqnarray}
	 \Sigma_1 (x) &=& \frac{2}{\cos^\beta i}(r_0\rho_c) \ln\left[\frac{1+\sqrt{y_1}}{\sqrt{1-y_1}}\right],\\
	 \Sigma_{2} (x) &=& \frac{2}{\cos^\beta i} \rho_c r_0 q^{\frac{1}{2}} \frac{1}{\sqrt{1-y_{2}}}\tan^{-1}\sqrt{\frac{y_2}{1-y_2}},\\
	\label{eq_msurfisosphere}
	 \Sigma_{3} (x) &=& \frac{2}{\cos^\beta  i} \rho_c r_0 q^{\frac{2}{3}}\frac{1}{1-y_3}\sqrt{y_3},\\
	 \Sigma_{4} (x) &=& \frac{2}{\cos^\beta i}\rho_c r_0 q^{\frac{3}{4}}\frac{1}{(1-y_4)^{3/2}}\frac{1}{2}\nonumber\\
			 &&\times \Bigg\{
			 	\sqrt{y_4}\sqrt{1-y_4} +  \tan^{-1}\sqrt{\frac{y_4}{1-y_4}}\Bigg\},
\end{eqnarray}
where $y_n = (1-x^2)(1-q^{2/n})$. The profiles of higher orders can be derived by applying successively the
integral transform Eq.~\ref{eq_integraltransform}.

The profile $n=4$ applies for isothermal self-gravitating pressurized cylinders. For cylinders exist a maximum 
mass line density given by $[M/l]_{\rm max}=2K/G$ where $G$ is the gravitational constant. 
$K$ is a constant given by $K= kT / (\mu m_{\rm H})$ where $T$ is the effective temperature, $k$ the Boltzmann
constant, $\mu$ the mean molecular weight and $m_{\rm H}$ is the atomic mass of hydrogen.
If we replace the pressure
ratio with $q = (1-f)^2$ where $f=(M/l)/[M/l]_{\rm max}\le 1$ is the normalized mass line density we obtain the
expression given in the work of \citet{Fischera2012b}. The profile for $n=3$ closely describes the profile of Bonnor-Ebert spheres
with overdensities less than $\sim 100$ \citep{Fischera2014b}. 

\subsection{\label{sect_gaussapprox}Gaussian approximation ($n\rightarrow \infty$)}

Under certain circumstances the inner region of the profile can be approximated by a Gaussian
function as will be shown in the following where the width is related to physical parameters as 
the overdensity $q^{-1}$ and the inner radius $r_0$.

The density profile can in general be expressed through 
\begin{equation}
	\rho(x) =  \rho_{\rm c} e^{-\frac{n}{2}\ln\left[1+(x r_{\rm cl}/r_0)^2\right]},
\end{equation}
where 
\begin{equation}
	\frac{r^2_{\rm cl}}{r^2_0} = q^{-\frac{2}{n}}\left(1-q^{\frac{2}{n}}\right).
\end{equation}
Where $(x r_{\rm cl}/r_0)^2\ll 1$
we can linearize the logarithm using $\ln [1+(x r_{\rm cl}/r_0)^2]\approx (x r_{\rm cl}/r_0)^2$ 
and we obtain a Gaussian density profile
\begin{equation}
	\rho(x) =  \rho_{\rm c} e^{-\frac{1}{2}\left(\frac{x}{\sigma_\rho}\right)^2},
\end{equation}
where the variance is given by
\begin{equation}
	\sigma_{\rho}^2 = \frac{1}{n}\frac{r^2_{0}}{r^2_{\rm cl}} = \frac{1}{n}\frac{q^{\frac{2}{n}}}{1-q^{\frac{2}{n}}}.
\end{equation}
The approximation improves with power $n$. In case of a pressure ratio $q$ 
the density profile becomes approximately a Gaussian 
function for all impact parameters if $n\gg-2\ln q / \ln 2$ or $q^{\frac{2}{n}}\gg \frac{1}{2}$.
For large $n$ the variance becomes
\begin{equation}
	\sigma_\rho^2 \rightarrow -\frac{1}{2 \ln q},
\end{equation}
which decreases slowly with overdensity.

In a similar approach we can derive the asymptotic profile of the   mass surface density. 
Considering the same condition for $n$ as above we obtain for example for
\begin{equation}
	(1-y_n)^{\frac{1-n}{2}} = q^{\frac{1-n}{n}} e^{-\frac{1}{2} \left(\frac{x}{\sigma_{\Sigma}}\right)^{2}},
\end{equation}
also a Gaussian form where the variance of the   mass surface density is given by
\begin{equation}
	\label{eq_gaussvariance}
	\sigma^2_{\Sigma} = \frac{1}{{n-1}}\frac{r_{0}^2}{r_{\rm cl}^2}=\frac{1}{n-1}\frac{q^{\frac{2}{n}}}{1-q^{\frac{2}{n}}},
\end{equation}
where $\sigma_{\Sigma}\approx \sigma_\rho=\sigma$ for large $n$.

The central region ($x\ll r_0/r_{\rm cl}$) of the profile is therefore approximately given by a Gaussian function
\begin{equation}
	\label{eq_gaussapprox}
	\Sigma_n \approx \rho_{\rm c}r_0
		e^{-\frac{1}{2}\left(\frac{x}{\sigma}\right)^2} {\rm B}\left(\frac{n-1}{2},\frac{1}{2}\right)\left(1-I_{q^{\frac{2}{n}}}\left(\frac{n-1}{2},\frac{1}{2}\right)\right).
\end{equation}
The approximation provides the exact central mass surface density. 

We want to consider the case of high overdensity and large power $n$ so that the
contribution of the incomplete beta function in the central region of the cloud 
becomes negligible (see Sect.~\ref{sect_incompletebeta}). 
In the limit of large $n$ the beta function becomes 
\begin{equation}
	\label{eq_betaapprox}
	\frac{\Gamma\left(\frac{n-1}{2}\right)\Gamma\left(\frac{1}{2}\right)}{\Gamma\left(\frac{n}{2}\right)}\rightarrow \sqrt{\frac{2\pi}{n}}.
\end{equation}
Replacing $n$ through the variance of the density profile we obtain for the asymptotic profile for given overdensity
\begin{equation}
	\label{eq_approxgauss}
	\Sigma_\infty(x) \rightarrow \rho_{\rm c} \sqrt{2\pi}\sigma r_{\rm cl}\,
		e^{-\frac{1}{2}\left(\frac{x}{\sigma}\right)^2} = \Sigma_\infty(0) e^{-\frac{1}{2}\left(\frac{x}{\sigma}\right)^2}.
\end{equation}

\section{\label{app_pdfasymptotes}Asymptotes of the PDF}

\subsection{\label{sect_pdfapproxsmall}Asymptotes in the limit $y_n\rightarrow 0$ (low $X_n$)}

In the limit of large impact parameters ($x\rightarrow 1$) 
it follows that $y_n\rightarrow 0$ and consequently $u_{\rm max} \approx \sqrt{y_n}\ll 1$.
The integrand in Eq.~\ref{eq_profile} is approximately a constant so that the unit-free mass surface density
becomes
\begin{equation}
	X_n \approx \sqrt{y_n}.
\end{equation}

From Eq.~\ref{eq_pdfsphere} we find directly the corresponding asymptotic behavior for spheres which is given by
\begin{equation}
	\label{eq_pdfsphapproxsmall}
	P_{\rm sph}(X_n)  \approx \frac{2}{1-q^{2/n}} X_n.
\end{equation}
For cylinders we find from Eq.~\ref{eq_pdfcylinder}
\begin{equation}
	P_{\rm cyl}(X_n)\approx P_{\rm cyl}^{(a)}(X_n) \approx \frac{1}{1-q^{2/n}}X_n.
\end{equation}
For $q^{2/n}\rightarrow 0$ we obtain {the asymptote of} the probability function of 
homogeneous spheres or cylinders.

\subsection{\label{sect_pdfapproxlarge}Asymptotes in the limit $y_n \rightarrow 1$ (high $X_n$)}

\subsubsection{Asymptotes for $n>1$}
In the limit of low impact parameter ($x\rightarrow 0$) and high overdensity 
($q\ll 1$) we have $y_n\rightarrow 1-q^{2/n}\sim 1$. 
The normalized   mass surface density is then approximately given by
\begin{equation}
	\label{eq_pdfasymptotelarge}
	X_n \sim  (1-y_n)^{\frac{1-n}{2}} \zeta_n\gg 1,
\end{equation}
where $\zeta_n=0.5 \,{\rm B}((n-1)/2,1/2)$. 
We can use this relation to replace $1-y_n$
to obtain an expression of the PDF as function of the mass surface density.

For the PDF of spheres we find that at high mass surface densities the PDF approaches asymptotically a power law given by
\begin{equation}
	\label{eq_approxlargesph}
	P_{\rm sph}(X_n) \sim \frac{1}{n-1}\frac{2}{1-q^{2/n}}
	X_n^{-\frac{n+1}{n-1}} \zeta_{n}^{\frac{2}{n-1}}.
\end{equation}
For the asymptotic PDF of cylinders we find
\begin{equation}
	\label{eq_approxlargecyl}
	P^{(a)}_{\rm cyl}(X_n) \sim \frac{1}{n-1}\frac{1}{1-q^{2/n}}
		X_n^{-\frac{n}{n-1}} \zeta_{n}^{\frac{1}{n-1}}.
\end{equation}


{Replacing $1-y_n$ in the Eq.~\ref{eq_pdfcylinder1} with the above expression for the mass surface density  
provides the asymptotic behavior of the PDF of cylinders at the pole given by
\begin{equation}
	P_{\rm cyl}(X_n) \sim \frac{1}{n-1}\frac{1}{\sqrt{1-q^{2/n}}} \frac{X_n^{-\frac{n}{n-1}}\zeta^{\frac{1}{n-1}}}
			{\sqrt{1-(X_n/X_n(0))^{\frac{2}{n-1}}}},
\end{equation}
where $X_n(0) = q^{\frac{1-n}{n}}\zeta_n$ is the central mass surface density. A power law is only established 
for cylinders with sufficiently high overpressure so that $(X_n/ X_n(0))^{2/(n-1)}\ll 1$. 
}


\subsubsection{Asymptotes for $n=1$}

In the limit of large $y_1$ the mass surface density behaves as
\begin{equation}
	X_1 \sim \ln\frac{2}{\sqrt{1-y_1}}.
\end{equation}
Replacing
\begin{equation}
	1-y_1 \approx 4 \,e^{-2 X_1}
\end{equation}
in Eq.~\ref{eq_pdfsphere} and in Eq.~\ref{eq_pdfcylinder} provides the
{asymptotes
\begin{equation}
	P_{\rm sph}(X_1) \approx \frac{2}{1-q^2}(1-y_n) \sim \frac{8}{1-q^2} e^{-2 X_1}
\end{equation}
for and }
\begin{equation}
	P^{(a)}_{\rm cyl}(X_1) \approx \frac{1}{1-q^2}\sqrt{1-y_n} \sim \frac{2}{1-q^2} e^{-X_1}.
\end{equation}
For the special case $n=1$ the PDF at high mass surface densities is therefore approximately
described by a simple exponential function.
The asymptotic behavior of the PDF for a cylinder including the region at the pole is
\begin{equation}
	P_{\rm cyl}(X_1) \approx \frac{2}{\sqrt{1-q^2}}\frac{e^{-X_1}}{\sqrt{1-e^{-2(X_1(0)-X_1)}}},
\end{equation}
where $X_1(0) = \ln (2 q^{-1})$.

\subsubsection{Asymptotes for $n<1$}

As pointed out in Sect.~\ref{sect_msurf_nlt1} for $n<1$ the   mass surface density has a maximum
possible value. In the limit $y_n\rightarrow 1$ we have
\begin{equation}
	 X_n  \approx \frac{1}{1-n}\sqrt{y_n}\le \frac{1}{1-n}.
\end{equation}
As can be shown we have $P_{\rm sph}(X_n)\rightarrow 0$ and $P_{\rm cyl}^{(a)}(X_n)\rightarrow 0$ for $X_n\rightarrow 1/(1-n)$.

\subsection{Asymptote of the PDF for high/low $n$}

\subsubsection{\label{sect_pdfgauss}Asymptote in the limit $n\rightarrow \infty$}

As we have seen in the previous section in the limit of high $n$ the power
law slope at high   mass surface densities approaches asymptotically $\alpha=1$.
The corresponding PDF can be directly obtained from Eq.~\ref{eq_approxlargesph}.
Replacing $n-1$ by the standard deviation of the Gaussian approximation 
we get for spheres with high overdensity in the limit of $n\rightarrow \infty$
\begin{equation}
	P_{\rm sph}(X_n) \sim 2 \sigma^2 X_{n}^{-1}.
\end{equation}
The same result is obtained from Eq.~\ref{eq_approxgauss} by deriving the
corresponding derivative and using Eq.~\ref{eq_probability}.
For the asymptotic PDF of cylinders Eq.~\ref{eq_approxlargecyl}
\begin{equation}
	P^{(a)}_{\rm cyl}(X_n) \sim \sigma^2 X_{n}^{-1}.
\end{equation}

\subsubsection{\label{sect_approx_n0}Asymptote in the limit $n\rightarrow 0$}

In the limit $n\rightarrow 0$ it follows from Eq.~\ref{eq_profile2} for the   mass surface density
\begin{equation}
	X_n \rightarrow \frac{1}{1-n} \sqrt{y_n}.
\end{equation}
In case of spheres the PDF of the mass surface density becomes 
\begin{equation}
	P_{\rm sph}(X_n) \approx \frac{2}{1-q^{2/n}}\sqrt{y_n}\rightarrow 2\sqrt{1-x^2},
\end{equation}
which is the PDF of homogeneous spheres.
Likewise, we find for the asymptotic PDF of cylindrical clouds in the limit $n\rightarrow 0$ that
\begin{equation}
	P^{(a)}_{\rm cyl}(X_n) \approx P_{\rm cyl}(X_n)\rightarrow \frac{\sqrt{1-x^2}}{x},
\end{equation}
which is the PDF of a homogeneous cylinder.

\section{\label{app_pdfmaxima}Condition for PDF maxima}

The maxima position were derived for both linear and logarithmic PDFs of spheres
and cylinders. For cylinders the asymptotic PDF
as defined in Eq.~\ref{eq_pdfcylinder} was considered. 

\subsection{For linear values ($P(\Sigma_n)$)}

The condition for maxima of the linear PDF is given by 
\begin{equation}
	\frac{{\rm d}P}{{\rm d}y_n}(X_n) = 0.
\end{equation}
This leads to
\begin{equation}
	X_n y_n^{\frac{3}{2}}(n^2-1)+ y_n (n+2) - 1 = 0
\end{equation}
in case spheres and to 
\begin{equation}
	X_n y_n^{\frac{3}{2}}n (n-1) +y_n (n+1)-1 = 0
\end{equation}
for cylinders.
For $n=1$ and $n=3$ the maxima are simple analytical expressions
listed in Tab.~\ref{table_pdfmaxima}.



%

\subsubsection{\label{app_maxasymptote}Approximation for $n\gg 1$}

In the limit $n\gg 1$
the condition for maxima of the linear PDF is equal for spheres and cylinders and is given by
\begin{equation}
	X_n y_n^{\frac{3}{2}} n^2 + y_n n -1 = 0.
\end{equation}
As $y_n\rightarrow 0$ for $n\rightarrow \infty$ it follows from Eq.~\ref{eq_sigmaapprox} and
Eq.~\ref{eq_betaapprox} that the mass surface density behaves approximately as 
\begin{equation}
	X_n \sim e^{\frac{n}{2}y_n}\frac{1}{2}\sqrt{\frac{2\pi}{n}} P(\chi^2,1),
\end{equation}
where $P(\chi^2,1)$ is the PDF of the $\chi^2$-distribution and where
\begin{equation}
	\chi^2 \rightarrow  n y_n.
\end{equation}
In the limit of $n\gg 1$ the condition for maxima becomes a function of $n y_n=C$ where $C$ is a constant.
Solving
\begin{equation}
	\label{eq_approx2}
	e^{C/2}\sqrt{\frac{\pi}{2}} P(C,1) C^{\frac{3}{2}}+C-1 = 0
\end{equation}
provides $C\approx 0.58404$. The maxima position is therefore approximately given by
\begin{eqnarray}
	[y_n]_{\rm max} &=& C/n, \\
	{[X_{n}]}_{\rm max} &=& e^{C/2}\sqrt{\frac{\pi}{2n}}P(C,1).
\end{eqnarray}

\subsection{For logarithmic values ($\Sigma_n P(\Sigma_n)$)}

The maxima of the PDF of logarithm values are given by the condition 
\begin{equation}
	\frac{{\rm d} }{{\rm d}y_n}[X_n P(X_n)] =0.
\end{equation}
%
This leads to 
\begin{equation}
	2X_n^2  y_n (n-1) - X_n \left((n-1) \sqrt{y_n}+\frac{1-3 y_n}{\sqrt{y_n}}\right) = 1
\end{equation}
in case of spheres and to
\begin{equation}
	X_n^2 y_n (n-1) - X_n\left((n-1)\sqrt{y_n}+\frac{1-2 y_n}{\sqrt{y_n}}\right) =1
\end{equation}
in case of cylinders.
For $n=3$ the maxima positions are again simple analytical expressions given in
Tab.~\ref{table_pdfmaxima}.


%

%

\section{\label{sect_incompletebeta}Asymptotic behavior of the incomplete beta function}

To derive the asymptotic behavior of the incomplete beta function in Eq.~\ref{eq_profile} for the
mass surface density in the limit $n\gg 1$ we consider the approximation (Eq.~26.5.20, of \citet{Abramowitz1972})
\begin{eqnarray}
	I_\xi(a,b) &\sim& 1 - P(\chi^2,\nu), \nonumber\\
			&=& \left(2^{\frac{\nu}{2}}\Gamma\left(\frac{\nu}{2}\right)\right)^{-1}
 		\int\limits_{\chi^2}^{\infty}{\rm d}t\,t^{\frac{\nu}{2}-1}\,e^{-\frac{t}{2}},
\end{eqnarray}
where $P(\chi^2,\nu)$ is the $\chi^2$ distribution function of $\nu$ events where
\begin{eqnarray}
	\chi^2 & = & (a+b-1)(1-\xi)(3-\xi) - (1-\xi)(b-1),\\
		 \nu&=&2 b.
\end{eqnarray}
In our case we have $a=(n-1)/2$ and $b=1/2$. The incomplete beta function is then 
approximately given by
\begin{equation}
	\label{eq_approxibeta}
	I_{1-y_n}\left(\frac{n-1}{2},\frac{1}{2}\right) \sim \frac{1}{\sqrt{2\pi}}\int\limits_{\chi^2}^{\infty}{\rm d}t\,t^{-\frac{1}{2}} e^{-\frac{t}{2}},
\end{equation}
where 
\begin{equation}
	\chi^2 =\left(\frac{n-2}{2}\right)y_n (2+y_n) + y_n\frac{1}{2}.
\end{equation}
For given power $n$ the approximation improves with increasing $\xi$.

In the limit $y_n\rightarrow 0$ we can use the replacement
\begin{equation}
	(1-y_n)^{\frac{1-n}{2}} \sim e^{\frac{n-1}{2}y_n},
\end{equation}
so that the   mass surface density becomes
\begin{equation}
	\label{eq_sigmaapprox}
	\Sigma_n \sim 2 r_0\rho_{\rm c} q^{\frac{n-1}{n}}e^{-\frac{1-n}{2}y_n}\frac{1}{2}{\rm B}\left(\frac{n-1}{2},\frac{1}{2}\right)P(\chi^2,1).
\end{equation}
In the limit $n\rightarrow \infty$ we have
\begin{equation}
	\chi^2 \approx n y_n \rightarrow - 2 \,(1-x^2)\,\ln q,
\end{equation}
so that the beta function becomes independent of $n$.

%

\subsection{Incomplete beta function for $b=1/2$}

The power law approximation of the PDF at large   mass surface
densities for $n>1$ as presented in Sect.~\ref{app_pdfasymptotes} are valid for negligible contribution
of the incomplete beta function to the   mass surface density. We have seen in Fig.~\ref{fig_pdf} that the
power law is only a good representation for large   mass surface densities and that the approximation of
the PDF improves for larger $n$. 

\begin{figure}[htbp]
	\includegraphics[width=\hsize]{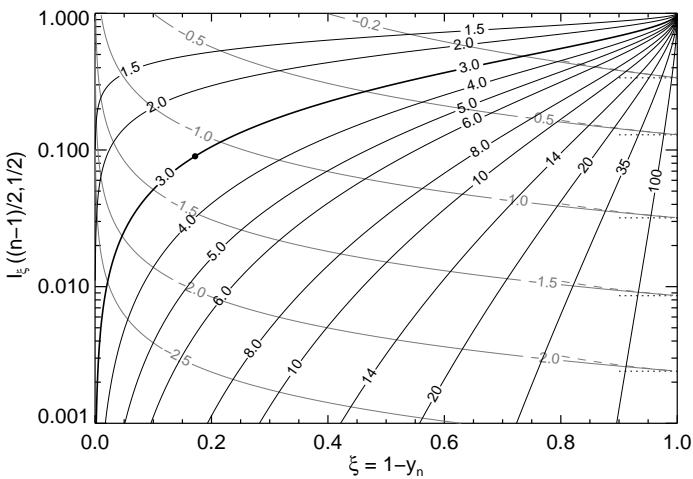}
	\caption{\label{fig_incompletebeta2} Incomplete beta function $I_\xi(a,b)$ (black contours)
	for $a=(n-1)/2$ and $b=1/2$ as function of $\xi=1-y_n$.
	The lines are labeled with the corresponding power $n$. The line for a power $n=3$
	is emphasized through a thick line. The gray lines
	correspond to $\xi=q^{2/n}$ for fixed density ratio $q$ and varying power $n$. The lines
	are labeled with $\log_{10} q$. The gray dashed lines are obtained using the approximation 
	of the incomplete beta function (Eq.~\ref{eq_approxibeta}). The asymptotic value of the incomplete
	beta function for given $q$ in the limit $n\rightarrow \infty$ is indicated through dotted lines.
	The filled circle corresponds to a power $n=3$ and an overdensity
	$q_{\rm crit}^{-1} = 14.04$ of a critical stable sphere.
	}
\end{figure}

Fig.~\ref{fig_incompletebeta2} shows the value of the incomplete beta function as given in Eq.~\ref{eq_profile}
for different assumptions for the powers $n$ and the density ratio $q$. As we see the value of the incomplete
beta function for given $q$ decreases for larger $n$. In the limit $n\rightarrow \infty$ for given $q$ we obtain the asymptotic
value of the incomplete beta function given in the previous section. In the limit $n\rightarrow 1$ we have
\begin{equation}
	I_{1-y_n}\left(\frac{n-1}{2},\frac{1}{2}\right) \rightarrow 1.
\end{equation}


\bibliographystyle{aa} 
\bibliography{fischerareference}

\end{document}